\documentclass[journal, 12pt, onecolumn,draftclsno
foot,]{IEEEtran}

\ifCLASSINFOpdf
   \usepackage[pdftex]{graphicx}
   \graphicspath{{../pdf/}{../jpeg/}}
   \DeclareGraphicsExtensions{.pdf,.jpeg,.png}
\else
   \usepackage[dvips]{graphicx}
   \graphicspath{{../eps/}}
   \DeclareGraphicsExtensions{.eps}
\fi

\usepackage{amsmath}
\usepackage{bm}
\usepackage{amssymb}
\usepackage{subfigure}
\usepackage{color}

\begin{document}

\title{Learning to Optimize Resource Assignment for Task Offloading in Mobile Edge Computing}

\author{Yurong~Qian,~\IEEEmembership{Student~Member,~IEEE}, Jindan~Xu,~\IEEEmembership{Member,~IEEE}, Shuhan~Zhu,~\IEEEmembership{Student~Member,~IEEE}, Wei~Xu,~\IEEEmembership{Senior~Member,~IEEE}, Lisheng Fan,~\IEEEmembership{Member,~IEEE}, and George K. Karagiannidis,~\IEEEmembership{Fellow,~IEEE}
\thanks{This work was supported in part by the National Key Research and Development Program 2020YFB1806600, the NSFC under grants 62022026 and 61871139, and the Natural Science Foundation of Jiangsu Province for Distinguished Young Scholars under Grant BK20190012. The research of G. K. Karagiannidis has been co-financed by the European Regional Development Fund by the European Union and Greek national funds through the Operational Program Competitiveness, Entrepreneurship and Innovation, under the call: ``Special Actions: Aquaculture--Industrial Materials--Open Innovation in Culture'' (project code: T6YBP-00134). The associate editor coordinating the review of this article and approving it for publication was Dr. Santiago Mazuelas. \emph{(Corresponding authors: Jindan Xu; Wei Xu.)}} 
\thanks{Y. Qian, J. Xu, S. Zhu, and W. Xu are with the National Mobile Communications Research Laboratory, Southeast University, Nanjing 210096, China. W. Xu is also with Purple Mountain Laboratories, Nanjing 211111 (e-mail: qianyr@seu.edu.cn; jdxu@seu.edu.cn; shzhu@seu.edu.cn; wxu@seu.edu.cn).}
\thanks{L. Fan is with the School of Computer Science, Guangzhou University, Guangzhou 510006, China (e-mail: lsfan@gzhu.edu.cn). }
\thanks{G. K. Karagiannidis is with the Electrical and Computer Engineering Department, Aristotle University of Thessaloniki, 54 124 Thessaloniki, Greece (e-mail: geokarag@auth.gr). }
} 
\maketitle

\begin{abstract}
 In this paper, we consider a multiuser mobile edge computing (MEC) system, where a mixed-integer offloading strategy is used to assist the resource assignment for task offloading. Although the conventional branch and bound (BnB) approach can be applied to solve this problem, a huge burden of computational complexity arises which limits the application of BnB. To address this issue, we propose an intelligent BnB (IBnB) approach which applies deep learning (DL) to learn the pruning strategy of the BnB approach. By using this learning scheme, the structure of the BnB approach ensures near-optimal performance and meanwhile DL-based pruning strategy significantly reduces the complexity. Numerical results verify that the proposed IBnB approach achieves optimal performance with complexity reduced by over $80\%$.

\end{abstract}

\IEEEpeerreviewmaketitle

\section{Introduction}
Smart mobile devices (MDs) are indispensable in modern daily life, whereas an increasing number of MDs has led to a huge challenge of handling a massive amount of data. Mobile edge computing (MEC) is an emerging architecture that can potentially address this challenge \cite{1}, through providing the ability of computation offloading to edge networks \cite{2}, which accelerates data computing, and improves user experience. 


Applying MEC, the MDs dynamically offload computation tasks to the edge computing access points (CAPs) \cite{5}. Solving the resource assignment problem of task offloading plays an essential role in the implementation of MEC, which ensures an efficient use of resources, low latency, and reduced energy consumption \cite{6}. Offloading assignment problem has been extensively studied recently \cite{7}-\cite{10}. In \cite{7}, the authors considered non-orthogonal multiple access (NOMA) transmission with offloaded workloads and proposed a distributed algorithm. 

The problem of resource assignment, however, is a combinatorial optimization problem, whose solution can be achieved optimally by exhaustive search methods \cite{9}, or suboptimally by, e.g., relaxation approaches of convex optimization \cite{10}. Specifically in \cite{10}, linear-based and semidefinite-based methods were proposed to optimize the offloading problem. The convex optimization based algorithms usually suffered from noticeable performance loss due to the relaxation of discrete assignment variables to continuous values. In order to approach the optimal performance, exhaustive search based algorithms, e.g., the branch and bound (BnB) approach \cite{9}, were proposed for MEC. However, exhaustive search methods are known to be computationally complex especially in a dense communication network with blooming mobile terminals.

Recently, deep learning (DL) has attracted much attention for its excellent performance in fitting arbitrary smooth functions \cite{12}-\cite{14}. In particular, the authors of \cite{12} used a deep neural network (DNN) to approximate the channel characteristics. In \cite{13}, the authors further made use of convolutional neural network (CNN) to learn a transmit power control strategy. 
However, these methods directly learnt the resource assignment solutions, which may lead to unsatisfied classification accuracy and insufficient generalization ability.




Motivated by the above, we propose an intelligent BnB (IBnB) approach to solve the problem of offloading resource assignment in a multiuser dynamic MEC system. The contributions of this paper are summarized as follows.
\begin{enumerate}[]
\item We propose an IBnB approach by making use of a joint model-driven structure of the conventional BnB and the data-driven DL technique. This proposed approach uses DL to learn the pruning strategy of BnB rather than directly learning the assignment solution by a block-box DNN, which improves the final prediction performance.
\item With the proposed IBnB approach, we further advice an adaptive threshold procedure which automatically adjusts the pruning threshold. Specifically, it enables automatically adjusting the threshold value when the initial threshold is too large to ensure a feasible solution. By using this adaptive scheme, the proposed approach enhances the model generalization capability and robustness.
\item Simulation results verify the superiority of the proposed approach. It reduces the complexity by nearly an order-of-magnitude while retaining the optimality of the solution in high probability.
\end{enumerate}


\section{System Model and Problem Formulation}

We consider a MEC system with time-varying channel consisting of a CAP and $S$ MDs, denoted by set \({\mathcal S} = \left\{ {1,2, \cdots, S} \right\}\), with insufficient resources. The MEC is applied in an orthogonal frequency division multiple access (OFDMA) scheme with time division duplexed (TDD) mode. There are $N$ time frames and $K$ orthogonal subchannels, denoted by \({\mathcal K} = \left\{ {1,2, \cdots, K} \right\}\). During each frame, all the tasks, coming up in the $S$ MDs, are suggested to be computed at the edge CAP through the $K$ subchannels. Let $\bf{X}$ be an indicator matrix of size $S \times K$ corresponding to this task offloading. The $(s, k)$th element of $\bf{X}$ is defined as 
\begin{equation}\label{offloading indicator}
{{[\bf{X}]}_{sk}} = \begin{cases}
1,  \emph{ \emph{if task $s$ is offloaded through subchannel $k$}},\\
0,  \emph{ \emph{otherwise}}.
\end{cases}
\end{equation}

Assuming that one subchannel can only handle the task from a single MD at a time, we have the constraint for $\bf{X}$ as
\begin{equation}\label{task allocation}
\sum\limits_{s \in {\cal S}} {{{[\bf{X}]}_{sk}}} \le 1.
\end{equation}

In addition, we consider a data partitioned oriented task model, e.g., virus scan and file/figure compression, where the tasks appear at MDs can be flexibly divided into several parts as subtasks \cite{15}. Define $L_{s}$ as the size of the task generated at MD $s$, and denoted by $l_{sk}$ as the size of one of the subtasks that is offloaded through subchannel $k$. Then, all the subtasks must be computed at the CAP, yielding,
\begin{equation}\label{task sapatation}
\sum\limits_{k \in {\cal K}} {{l_{sk}}}  = L_s.
\end{equation}

The parameter $l_{sk}$ can be reshaped as a matrix $\bf{L}$ of size $S \times K$ and $\left[\bf{L} \right]_{sk}= l_{sk}$. In general, we have $l_{sk}\le L_s$ where the equality $l_{sk}=L_s$ occurs when the conveyed subtasks through subchannel $k$ happens to be the entire task. 

In applications, latency and energy consumption are two key criteria in MEC networks \cite{5}. Typically, the offloading resource assignment can be formulated in the following problem with the objective function 
\begin{equation}\label{optimization function}
\mathop {\min }\limits_{\bf{X}, \bf{L}}{\kern 1pt} {\kern 1pt}  \Psi \left( {\bf{X},\bf{L}} \right) \triangleq \lambda_t T \left( {\bf{X},\bf{L}} \right) + \lambda_e E\left({\bf{X},\bf{L}} \right),
\end{equation}
where $\lambda_t$ and $\lambda_e$ are the weights to balance the importance between latency, $T \left( {\bf{X},\bf{L}} \right)$, and energy consumption, $E\left({\bf{X},\bf{L}} \right)$. The two weights can be determined by the specific systems status, e.g., remaining battery life and tolerable latency.

As we consider the problem of offloading resource assignment, the system latency $T\left({\bf{X},\bf{L}} \right)$ in (\ref{optimization function}) is the transmission latency during offloading. Then, the transmission latency of subchannel $k$ due to the offloading is calculated as 
\begin{equation}\label{adaptive latency}
{t_k} = \sum\limits_{s \in {\cal S}}\frac{{{[\bf{X}]}_{sk}l_{sk}}}{{R_{sk}}},
\end{equation}
where $R_{sk}$ represents the uplink data rate of MD $s$ over the $k$th subchannel, which is defined as, 
\begin{equation}\label{energy definition}
R_{sk} = B\log_2\left(1+\frac{P_sh_{sk}}{N_0}\right),
\end{equation}
where $B$ is the communication bandwidth of the subchannel, $P_s$ is the transmit power of the $s$th MD, $N_0$ is the variance of the zero-mean additive white Gaussian noise (AWGN), and $h_{sk}$ is the channel gain between the edge CAP and the $s$th MD through the $k$th subchannel. Without loss of generality, the channel gains are assumed to be block-independent, i.e., invariant within one time frame and varying independently from one frame to another.

Considering the latency of the entire procedure of MEC, all the subchannels transmit their corresponding subtask data simultaneously. Then, the latency is in fact determined by the maximum value among all ${t_k}^{'s}$. It follows
\begin{equation}\label{latency definition}
T({\bf{X}},{\bf{L}}) = \max\limits_{k\in{\cal K}}\ t_k.
\end{equation}

The energy consumption $E\left({\bf{X},\bf{L}} \right)$ in (\ref{optimization function}) is defined as the sum of consumed energy of all the serving MDs for task data transmission. Accordingly, the energy consumption can be evaluated as
\begin{equation}\label{energy definition}
E({\bf{X}},{\bf{L}}) = \sum\limits_{k \in {\cal K}}\sum\limits_{s \in {\cal S}} P_s\frac{{{[\bf{X}]}_{sk}l_{sk}}}{{R_{sk}}},
\end{equation}


Considering equations (\ref{offloading indicator}), (\ref{task allocation}), (\ref{task sapatation}), (\ref{latency definition}), and (\ref{energy definition}), the optimization problem of the offloading resource assignment is a mixed integer optimization problem formulated as
\begin{subequations}\label{relaxed problem}
\begin{align}
\mathop {\min }\limits_{\bf{X}, \bf{L}}{\kern 1pt} {\kern 1pt}  &\Psi \left( {\bf{X},\bf{L}} \right) = \lambda_t\max\limits_{k\in{\cal K}}\ \left(\sum\limits_{s \in {\cal S}}\frac{{{[\bf{X}]}_{sk}l_{sk}}}{{R_{sk}}} \right)\nonumber\\
&\ \ \ \ \ \ \ \ \ \ \  + \lambda_e P_s\sum\limits_{k \in {\cal K}}\sum\limits_{s \in {\cal S}}\frac{{{[\bf{X}]}_{sk}l_{sk}}}{{R_{sk}}}\\
{\rm{s.t.}}\  &\sum\limits_{s \in {\cal S}} {{{[\bf{X}]}_{sk}}} \le  1, \forall k\in {\mathcal K} \\
&\sum\limits_{k \in {\cal K}} {{l_{sk}}}  = L_s, \forall s\in {\mathcal S}\\ 
&0 \le l_{sk} \le L_s, \forall k\in {\mathcal K},\ \forall s\in {\mathcal S}\\
&{{{[\bf{X}]}_{sk}}}\in\{0,1\}, \forall k\in {\mathcal K},\ \forall s\in {\mathcal S}.
\end{align}
\end{subequations}

\section{The Proposed IBnB Approach}


Note that the problem in (\ref{relaxed problem}) is a mixed integer nonlinear programming (MINLP) problem. To solve this problem, we propose a DL-based IBnB approach. By intelligently learning the sample data, the trained DNN helps determine the future resource assignment strategies, which significantly reduces complexity and ensures near-optimal performance.
\begin{figure*}[!t]
\centering
\subfigure[ Training stage]{
\includegraphics[height=1.9in]{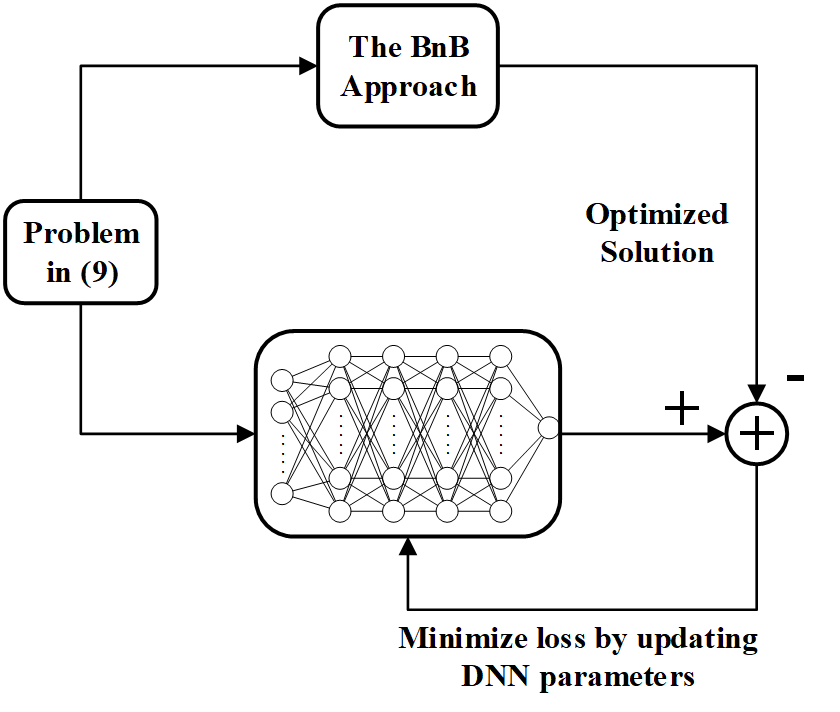} 
}
\subfigure[The DL-based branching process of the IBnB approach]{
\includegraphics[height=1.9in]{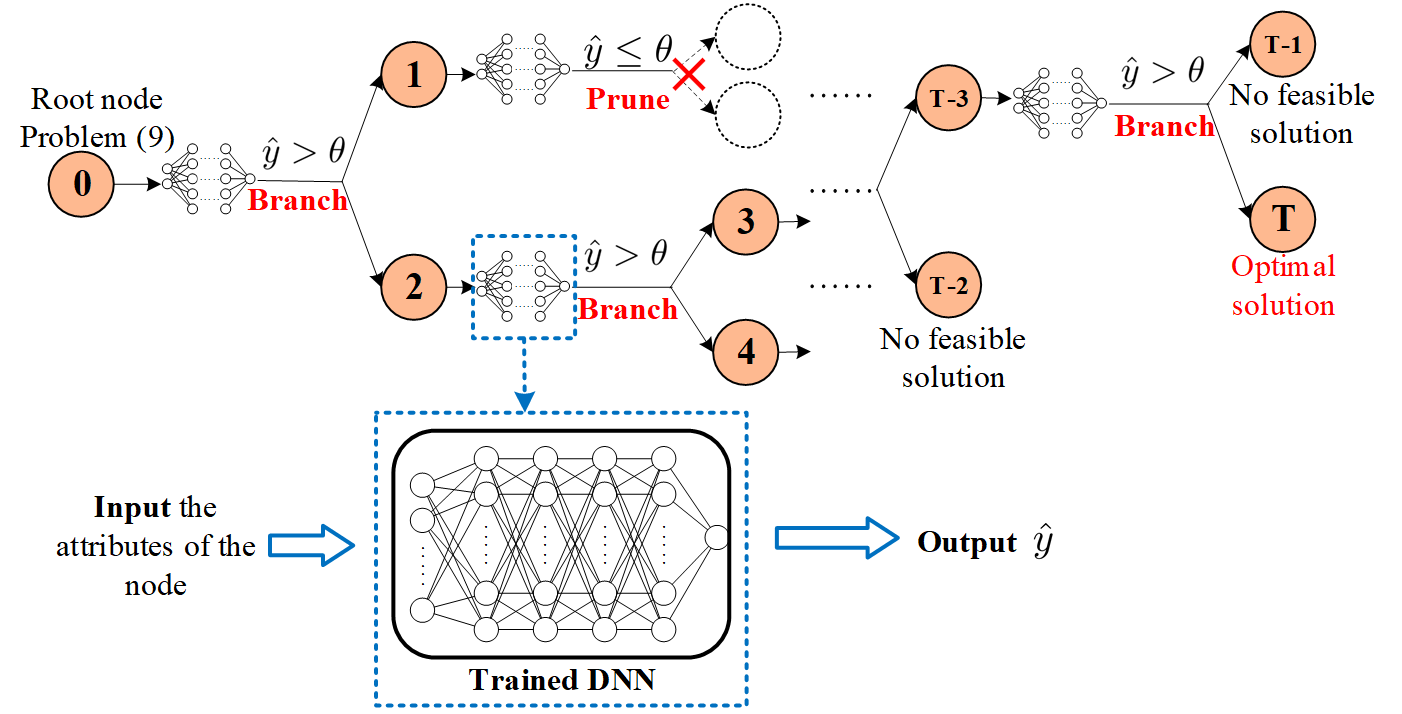} 
}
\caption{Schematic diagram of the proposed IBnB approach. }
\label{fig:1}
\end{figure*}

\subsection{Review of Conventional BnB Approach}
The BnB approach is based on an exhaustive search mechanism, which systematically enumerates all feasible solutions by traversing the BnB search tree. Specifically, the original optimization problem in (\ref{relaxed problem}) is regarded as the root node of the BnB tree, and then branching and bounding are implemented to construct and search the tree. \emph{Branching} is the process of dividing a big parent problem into two subproblems by adding mutually exclusive and complete constraints. This process is regarded as adding child nodes to the tree continuously. \emph{Bounding} is the process of solving and checking the upper and lower bounds of the subproblems in the process of branching. When the subproblem produces a better solution than the current bound, the current node is reserved to branch. Otherwise, this node is pruned.
\begin{table}[!t] \normalsize
\centering
\begin{tabular}{r l}
\hline
\multicolumn{2}{l}{{\textbf {Algorithm 1}} The BnB Approach}\\
\hline
1&{\textbf {Initialize:}} Set of the tree nodes:  $\mathcal N = \{N_0\}$, Upper bound of the objective value\\ & $\Psi$ in (9a): $Z_{\rm{UB}} = \infty$\\ 
2&{\textbf {While}} $\mathcal N \neq \emptyset$ {\textbf {do}}\\
3&\ \ $j$ $\leftarrow$ $j+1$;\\
4&\ \ $N_{j}$ $\leftarrow$ pop the first node from $\mathcal N$ ;\\
5&\ \ Without considering the integer constraints in (9e), calculate $(\textbf{\emph{x}}^{(j)}, \textbf{\emph{l}}^{(j)}, \Psi ^{(j)})$ \\ &\ \ using (9a), s.t. (9b), (9c), (9d), and the special constraint ${\mathcal C}_t$ of this node $N_{j}$;\\
6& \ \  {\textbf {if}} $\textbf{\emph{x}}^{(j)}$ is not integer {\textbf {then}}\\ 
7& \ \ \ \  {\textbf {if}} $\Psi^{(j)}\le Z_{\rm{UB}}$ {\textbf {then}}\\ 
8& \ \ \ \ \ \  Find the first non-integer component $x_i^{(j)}$ in $\textbf{\emph{x}}^{(j)}$; \\
9& \ \ \ \ \ \  Branch the node into two child nodes $N_{j}^{(1)}$, $N_{j}^{(2)}$ by adding two \\ &\ \ \ \ \ \   constraints ${\mathcal C}_j^{(1)}$, ${\mathcal C}_j^{(2)}$ respectively,\\ &\ \ \ \ \ \   ${\mathcal C}_j^{(1)}$: ${\mathcal C}_j$ $\cup$ $\{x_{i} \le \lfloor x_{i}^{(j)} \rfloor\}$, ${\mathcal C}_j^{(2)}$: ${\mathcal C}_j$ $\cup$ $\{x_{i} \ge \lfloor x_{i}^{(j)} \rfloor +1\}$;\\
10&\ \ \ \ \ \  $\mathcal N$ $\leftarrow$ $\mathcal N \cup \{N_{j}^{(1)}, N_{j}^{(2)}\}$;\\
11& \ \ \ \ {\textbf {end if}}\\
12&\ \ {\textbf {else if}} $\Psi^{(j)}< Z_{\rm{UB}}$\\
13&\ \ \ \ $Z_{\rm{UB}} \leftarrow \Psi^{(j)}$;\\
14&\ \ \ \ $\textbf{\emph{x}}^{*}, \textbf{\emph{l}}^{*}\leftarrow\textbf{\emph{x}}^{(j)}, \textbf{\emph{l}}^{(j)}$;\\
15& \ \ {\textbf {end if}}\\
16& {\textbf {end while}}\\
\hline
\end{tabular}
 \end{table}

Algorithm 1 exemplifies the procedure of applying the BnB approach, where we vectorize the solution matrices $\bf{X}$ and $\bf{L}$ in (\ref{relaxed problem}) as equivalent vector variables $\textbf{\emph{x}}=( x_1, x_2, \cdots, x_{S\times K} )^\mathrm{T}$ and $\textbf{\emph{l}}=( l_1, l_2, \cdots, l_{S\times K} )^\mathrm{T}$. Denote $\textbf{\emph{x}}^{(j)}=( x_1^{(j)}, x_2^{(j)}, \cdots, x_{S\times K}^{(j)} )^\mathrm{T}$ and $\textbf{\emph{l}}^{(j)}=( l_1^{(j)}, l_2^{(j)}, \cdots, l_{S\times K}^{(j)} )^\mathrm{T}$ as the optimal relaxation solutions to the subproblem of the $j$th node, and $\Psi ^{(j)}$ is the corresponding objective value in (9a), where $j\in \{0,1,\cdots,J\}$, and $J$ is the number of nodes of the BnB search tree. 

 As the number of nodes in the BnB search tree is the number of iterations to solve the MINLP problem in (\ref{relaxed problem}), the complexity of the BnB approach is directly determined by the number of searched nodes. However, in the BnB search tree, only a few branches point to the optimal solution. A large number of redundant nodes lead to a high computational complexity. A better pruning strategy can make it possible to find the optimal solution with a lower complexity. In the next subsection, we introduce a DL-based method to learn an intelligent pruning strategy, which achieves the optimal performance with significantly reduced complexity.


\subsection{The Proposed Low-Complexity IBnB Approach}
Based on recent researches, DL performs well in solving some NP-hard and nonconvex problems \cite{12}-\cite{14}. However, for combinational problems with binary, nonbinary and continuous variables, like the problem in (\ref{relaxed problem}), a direct application of DL can hardly achieve a satisfactory solution. Moreover, DL usually requires a large amount of training data, which is a huge challenge for mobile communication applications especially with MEC \cite{17}. 

To solve the above problems, we propose a low-complexity IBnB approach. Specifically, we design a novel pruning strategy by applying DL, which avoids branching (almost) all nodes as in the conventional BnB approach. We use a DNN to approximate the unknown mapping between the attributes of BnB tree nodes and the pruning decisions in the BnB search tree. In particular, the proposed DNN consists of 6 layers, which respectively has $\{m, 256, 256, 256, 256, 1\}$ neurons in each layer, where $m$ is the input dimension of the training sample data. At each layer except for the last one, the hyperbolic tangent function, tanh $f(x)=(e^x-e^{-x})/(e^x+e^{-x})$, is used as the activation function. In the last layer, we use the sigmoid function, $f(x)=1/(1+e^{-x})$, to map the output to the interval $(0,1)$. The Adam algorithm is chosen as the optimizer in the training phase, and the cross-entropy is used as the loss function. 

To train the DNN, we generate the dataset through the conventional BnB approach, denoted by ${\mathcal D}=\{{\mathcal I}, {\mathcal O}\}$, where ${\mathcal I}$ is the input dataset and ${\mathcal O}$ is the output dataset. The input data is the attributes of the node, including the variables $j$, $g$, $\textbf{\emph{x}}^{(j)}$, $\textbf{\emph{l}}^{(j)}$, $\Psi ^{(j)}$, and $f$, where $j$ is the node number, $g$ is the level of the node in the tree, $\textbf{\emph{x}}^{(j)}$ and $\textbf{\emph{l}}^{(j)}$ are the relaxation solutions of the subproblem of this node, $\Psi ^{(j)}$ is the objective value corresponding to the solutions, and $f$ is the flag indicating whether the subproblem is solvable. The output data in ${\mathcal O}$ is a binary variable $\{0,1\}$ indicating whether the input node is a parent node of the node corresponding to the optimal solution. If it is true, the output is 1, which means that the node is reserved to branch. Otherwise it is 0, which means that the node needs to be pruned. Fig. 1(a) depicts a diagram of the training procedure. 

The training process is periodically performed offline to update the parameters of the DNN. The trained DNN serves as a classifier in the process of the branching. The input of the trained DNN is the attributes of the node, while the output of the trained DNN is a scalar variable ranging in the interval $(0, 1)$, which is used to indicate the pruning strategy. Denote by $\hat y$ as the output of the trained DNN. A threshold, denoted by $\theta$, is set to distinguish the pruning decision $\rho$. By comparing the values of $\hat y$ and $\theta$, we set
\begin{equation}\label{threshold decision}
\rho= \begin{cases}
1,\ \emph{ \emph{if $\hat y > \theta$}},\\
0,\ \emph{ \emph{otherwise}}.
\end{cases} 
\end{equation}
Note that $\rho = 1$ implies that the current node is decided to branch into two child nodes, otherwise $\rho = 0$ corresponds to a pruned node. Fig. 1(b) elaborates the DL-based branching process with an example.
\begin{table}[!t] \normalsize
\centering
\begin{tabular}{r l}
\hline
\multicolumn{2}{l}{{\textbf {Algorithm 2}}  The Proposed IBnB Approach}\\
\hline
1&{\textbf {Initialize:}} $\mathcal N = \{N_0\}$, $\theta = \theta_0$, $Z_{\rm{UB}} = \infty$\\
2&{\textbf {While}} $Z_{\rm{UB}} = \infty$ {\textbf {do}}\\
3&\ \  {\textbf {While}} $\mathcal N \neq \emptyset$ {\textbf {do}}\\
4&\ \ \ \ Same as \textbf{Step 3-5} of {\textbf {Algorithm 1}};\\
5& \ \ \ \ {\textbf {if}} $\textbf{\emph{x}}^{(j)}$ is not integer {\textbf {then}}\\ 
6& \ \ \ \ \ \ {\textbf {if}} $\Psi^{(j)} < Z_{\rm{UB}}$ {\textbf {then}}\\ 
7& \ \ \ \ \ \ \ \ \textbf {Input} $j, g, f, \textbf{\emph{x}}^{(j)}, \textbf{\emph{l}}^{(j)}, \Psi ^{(j)}$ to the trained DNN; \\
8& \ \ \ \ \ \ \ \ \textbf {Output} $\hat y$;\\
9&\ \ \ \ \ \ \ \ Calculate $ \rho$ using (\ref{threshold decision});\\
10& \ \ \ \ \ \ \ \ {\textbf {if}} $\rho = 1$ {\textbf {then}}\\ 
11& \ \ \ \ \ \ \ \ \ \ Same as \textbf{Step 8-9} of {\textbf {Algorithm 1}};\\
12& \ \ \ \ \ \ \ \ {\textbf {end if}}\\
13& \ \ \ \ \ \ {\textbf {end if}}\\
14&\ \ \ \ {\textbf {else if}} $\Psi^{(j)}< Z_{\rm{UB}}$\\
15&\ \ \ \ \ \ $Z_{\rm{UB}} \leftarrow \Psi^{(j)}$;\\
16&\ \ \ \ \ \ $\textbf{\emph{x}}^{*}, \textbf{\emph{l}}^{*}\leftarrow\textbf{\emph{x}}^{(j)}, \textbf{\emph{l}}^{(j)}$;\\
17& \ \ \ \ {\textbf {end if}}\\
18& \ \ {\textbf {end while}}\\
19& \ \  $\theta_0$ $\leftarrow$  $\theta_0 \times \Delta \theta$;\\
20& {\textbf {end while}}\\
\hline
\end{tabular}
 \end{table}

Obviously, the threshold $\theta$ has an impact on the performance of the proposed IBnB approach. A lower threshold results in more nodes to be searched, which lifts the computation complexity, while a higher threshold may prune the branch with the optimal solution. By choosing an appropriate $\theta$, we flexibly reduce the number of searched nodes and guarantee the optimal performance of the proposed IBnB. It is clear that the best case is to prune all the nodes that do not lead to the optimal solution. However, an extreme case is over-pruning, which can lead to no feasible solution. Another extreme case is that a bitty threshold prunes few nodes, and has almost the same complexity as the conventional BnB.

Here, we design an adaptive threshold procedure to guarantee a feasible solution. This procedure enables to adjust the threshold automatically when the initial value of the threshold is too large to ensure a feasible solution for the optimization problem in (\ref{relaxed problem}). Specifically, an initial threshold is set, which in most cases can find the optimal solution but prune redundant nodes as much as possible. When the initial threshold is too large, excessive nodes are pruned and there can be no feasible solution left. To fix this, the threshold is lifted by a small step, i.e., $\theta \leftarrow \theta \cdot \Delta\theta$, and then those search trace back a layer to check the IBnB tree to find a solution.

The IBnB algorithm is summarized in \mbox{Algorithm 2}. In steps 4 and 11, the proposed IBnB performs the same steps to solve the relaxation problem of (\ref{relaxed problem}) and branch as the conventional BnB. In steps 7-9, the trained DNN is used to prune some redundant nodes with an appropriate threshold $\theta$, thereby reducing the number of searched nodes to significantly reduce complexity. Step 19 shows the process of adaptive threshold adjustment. The threshold adjustment and learning procedure repeats until a feasible solution is reached. 

\subsection{Complexity Analysis}
As shown in Section III-A, as long as there are enough iterations, the optimal solution to (\ref{relaxed problem}) can be found. As the training process is usually performed offline to update the parameters of the DNN, the structure and training process of DNN have a marginal impact on the complexity of the proposed approach when it is used online. In both the conventional BnB and the proposed IBnB approaches, the complexity comes from the process of repeatedly solving the MINLP problem in (\ref{relaxed problem}). Denote by ${\mathcal O} (S)$ as the complexity of solving the objective function once during the iteration. Then, the complexity of the conventional BnB is ${\mathcal O} (S\times J) $, and with an increasing number of MDs, $J$ will grow exponentially. Denote $T$ as the number of nodes of the IBnB search tree, so the complexity of the proposed IBnB is ${\mathcal O} (S\times T)$. By setting an appropriate threshold, the IBnB approach prunes most redundant nodes, and $T$ becomes proportional to the number of MDs. Obviously, the IBnB approach prune a large number of redundant nodes through an intelligent pruning strategy to greatly reduce the complexity.
\section{Experimental Results}
\begin{figure}[!t]
\centering
\begin{minipage}{1\textwidth}
\centering
\includegraphics[width=4in]{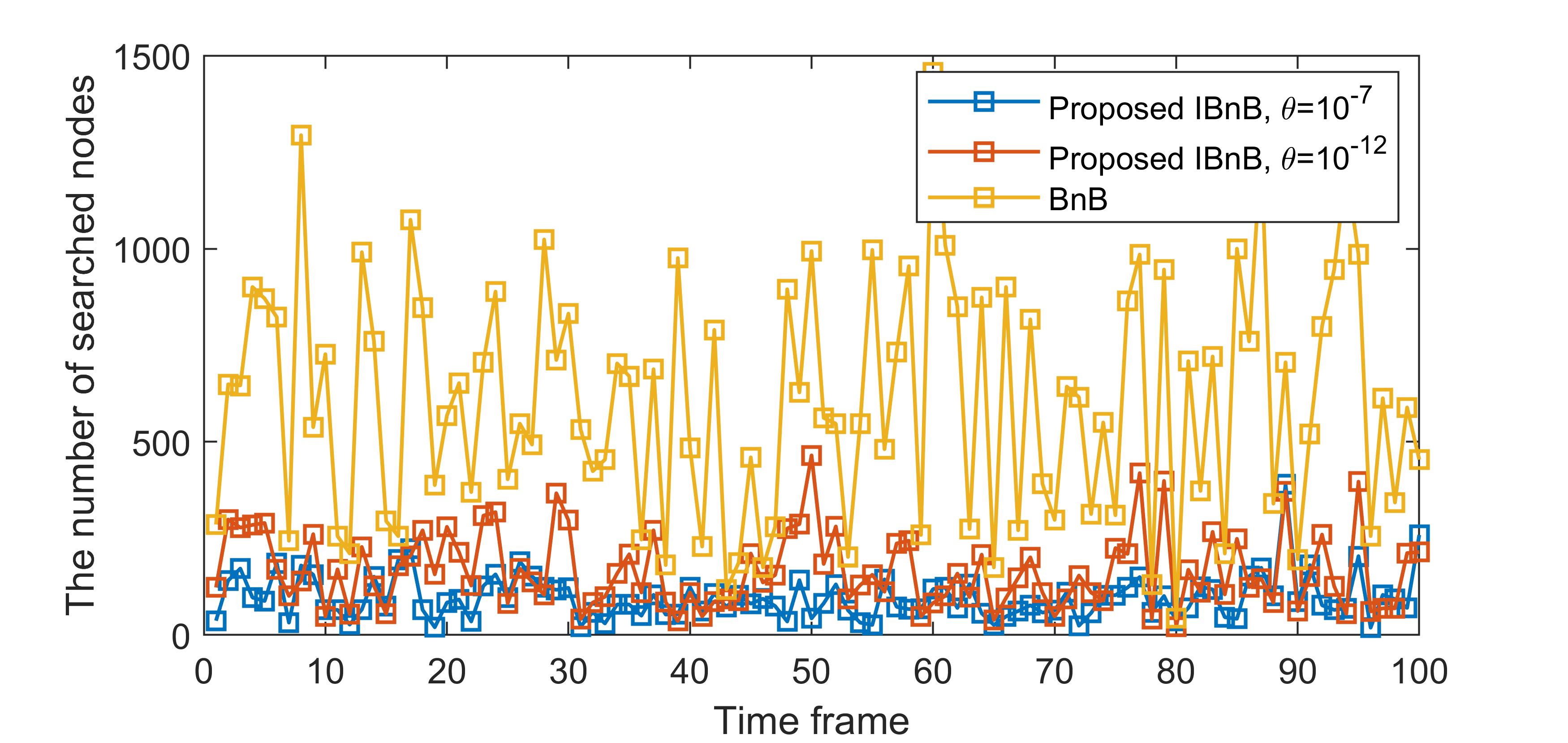}
\caption{Comparison of the number of searched nodes of the BnB and the proposed IBnB algorthms in different time frames.}
\end{minipage}
\begin{minipage}{1\textwidth}
\centering
\includegraphics[width=4in]{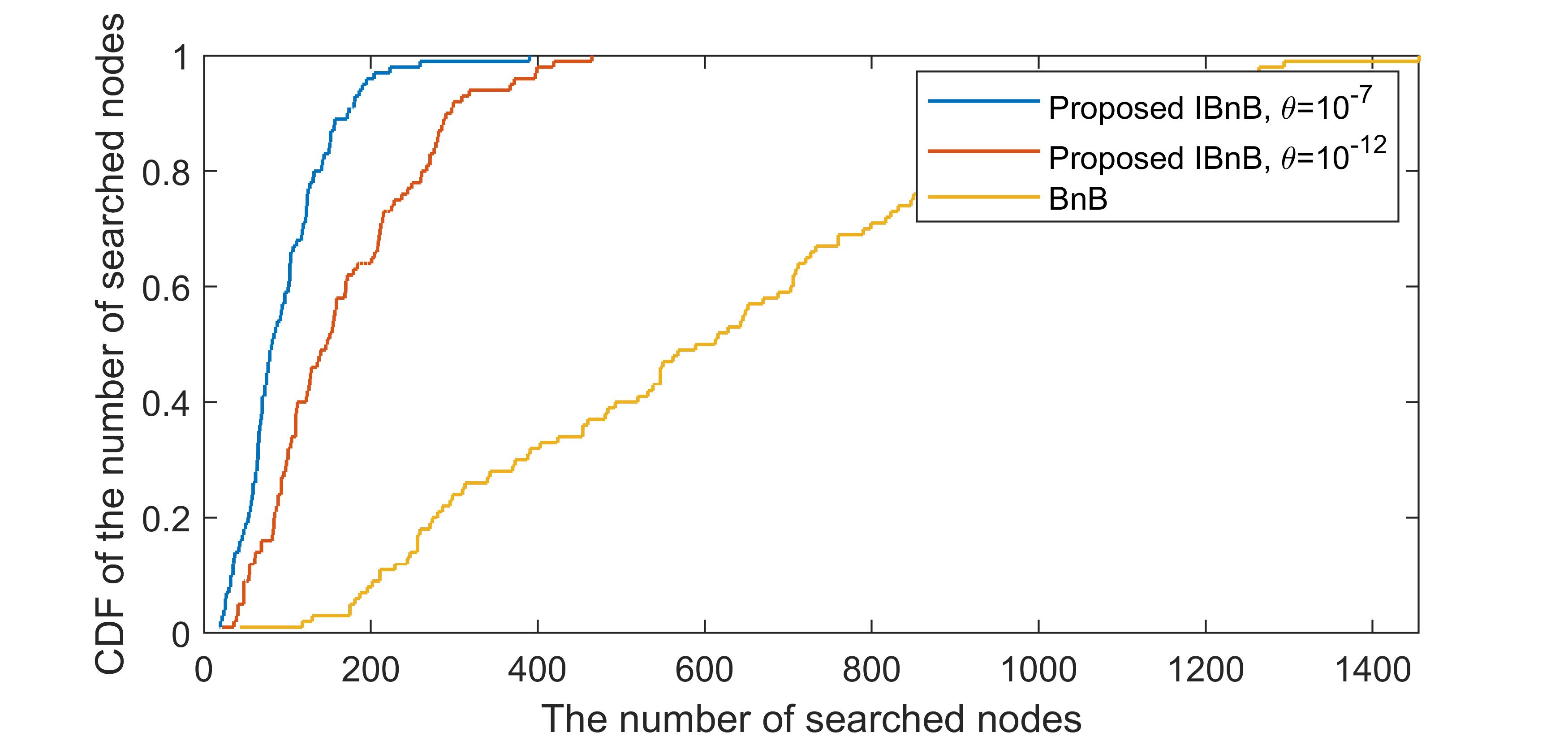}
\caption{Complexity comparison.}
\end{minipage}
\begin{minipage}{1\textwidth}
\centering
\includegraphics[width=4in]{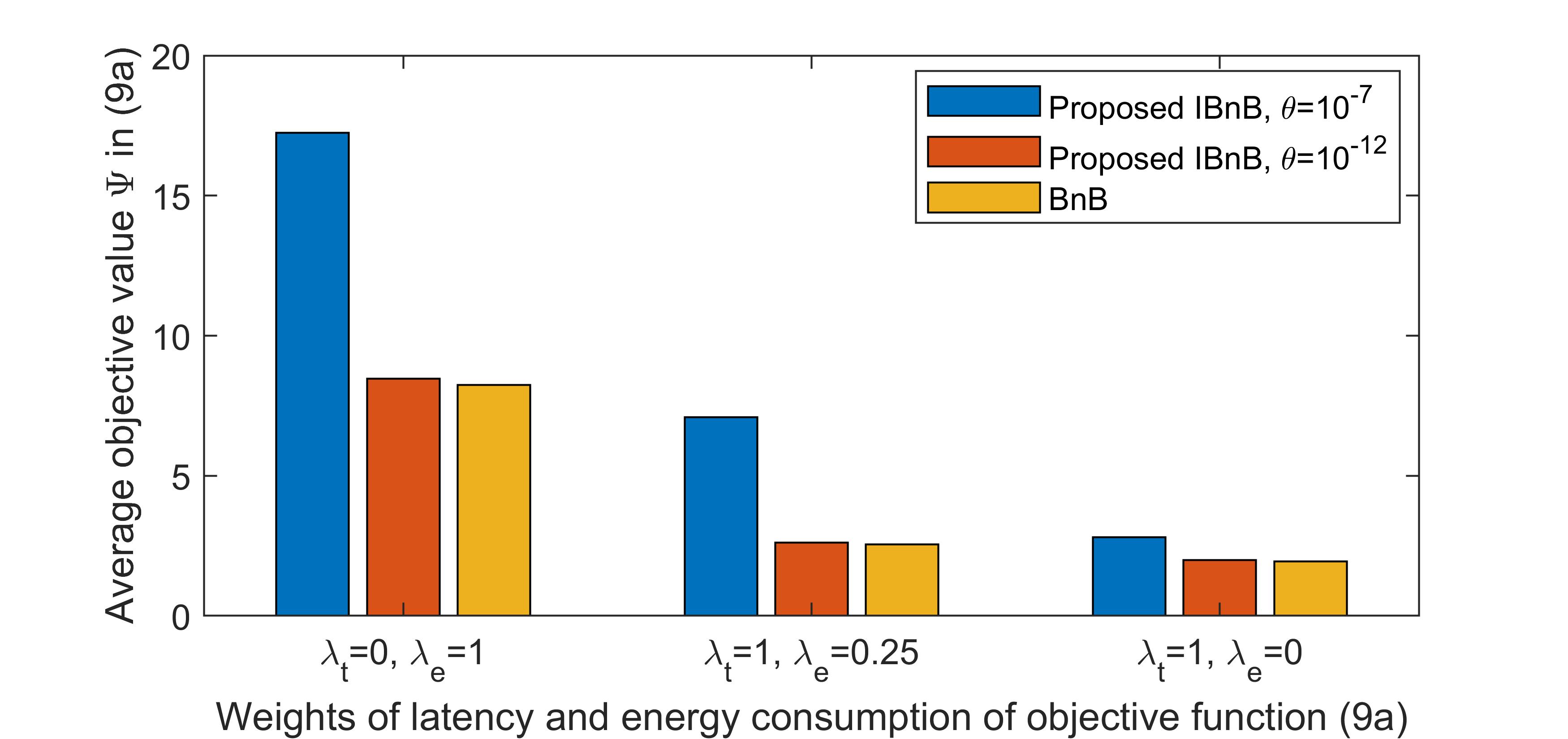}
\caption{Performance comparison of the BnB and the proposed IBnB algorthms under different weights of latency and energy consumption.}
\end{minipage}
\end{figure}
This section validates the efficiency of the proposed IBnB approach through simulations. In the simulations, the transmit power $P_s$ is chosen from a uniform distribution in the range of $[1.0, 1.5]$ W. The communication bandwidth $B$ is $10$ MHz. The channel gain $h_{sk}$ is generated from a Rayleigh fading channel model as in \cite{16}, and $N_0$ is set to $-110$ dBm. The CPU frequency of the CAP is $r = 2 \times {10^9}$ cycles/sec. The weight $\lambda_t$ is set to $1$, and $\lambda_e$ is set to $0.25$. Two initial thresholds are set to $\theta=10^{-7}$ and $\theta=10^{-12}$, and the adaptive threshold step parameter is set to $\Delta\theta=10^{-5}$. 


As the number of searched nodes in the BnB approach directly affects the complexity of the approach, we compare the number of searched nodes to reflect the complexity of the two approaches in Fig. 2. It shows that the complexity of the proposed IBnB is much less than that of the conventional BnB in all time frames. Moreover, from Fig. 2, it is verified that the variance of the complexity of our proposed IBnB is also greatly reduced compared to the conventional BnB. This signifies that the proposed IBnB is robust and appropriate for the delay-turbulance-sensitive tasks.

In Fig. 3, we provide the cumulative distribution function (CDF) of the number of searched nodes for the conventional BnB and proposed IBnB under different values of $\theta$. From \mbox{Fig. 3}, the proposed IBnB with the threshold of $10^{-7}$ can reduce the complexity of the conventional BnB to about $10\%$, while the proposed IBnB with the threshold of $10^{-12}$ can reduce the complexity to about $20\%$. Hence, we can conclude that the proposed IBnB is computationally efficient.


The average cost of the objective value $\Psi$ reflects the performance of the two approaches. In Fig. 4, we compare the performance of the two approaches under different weights of latency and energy consumption of (9a). The different weights may suit for different application scenarios. We can find from Fig. 4 that the proposed IBnB with the threshold of $10^{-12}$ approaches $98\%$ of the optimal performance by the conventional BnB, while at the complexity of only $20\%$. When the threshold is $10^{-7}$, the approach can also approach $80\%$ of the optimal performance. The experimental results show that our proposed approach improves the model’s generalization capability under different scenarios.



By adaptively adjusting the threshold, we can always find a feasible solution. In fact, if the initial threshold is low, e.g., lower than $10^{-12}$, the number of searched nodes becomes high and the complexity increases. On the other hand, if the initial threshold is high, e.g., higher than $10^{-7}$, the probability of achieving the optimal solution, or a feasible solution, decreases. After setting the initial threshold based on specific scenario parameters, e.g., the parameters of variable channels and the size of the offloading tasks, the trained DNN is in general able to find a feasible solution approaching the optimal performance, consuming less computation resource.

\section{Conclusion}
In this paper, we presented a model-and-data-driven offloading resource assignment approach in a MEC system. The proposed IBnB approach learnt the pruning strategy of the decision-making tree to significantly reduce the complexity. Simulation results were demonstrated to validate that the performance of the proposed IBnB is very close to the optimal one, and the complexity is only one-fifth or even lower compared to that of the conventional BnB.


\vspace{-0.3cm}

\begin{thebibliography}{1}

\bibitem{1}
H. Li, G. Shou, Y. Hu, and Z. Guo, ``Mobile edge computing: Progress and challenges," in {\emph {Proc. IEEE Int. Conf. Mobile Cloud Comput Services Eng. (MobileCloud)}}, pp. 83--84, 2016.
\bibitem{2}
S. S. D. Ali, H. Zhao, and H. Kim, ``Mobile edge computing: A promising paradigm for future communication systems," in {\emph {Proc. IEEE Region 10 Conf.}}, pp. 1183--1187, 2018.


\bibitem{5}
B. Dab, N. Aitsaadi, and R. Langar, ``A novel joint offloading and resource allocation scheme for mobile edge computing," in {\emph {Proc. 16th IEEE Annu. Consum. Commun. Netw. Conf. (CCNC)}}, pp. 1--2, Jan. 2019.
\bibitem{6}
S. Guo, B. Xiao, Y. Yang, and Y. Yang, ``Energy-efficient dynamic offloading and resource scheduling in mobile cloud computing," in {\emph {Proc}}. {\emph {IEEE INFOCOM}}, pp. 1--9, Apr. 2016.
\bibitem{7}
Y. Wu, K. Ni, C. Zhang, L. P. Qian, and D. H. K. Tsang, ``NOMA-assisted multi-access mobile edge computing: A joint optimization of computation offloading and time allocation," {\emph {IEEE Trans. Veh. Technol.}}, vol.~67, no.~12, pp.~12244--12258, Dec. 2018.
\bibitem{9}
G. Yang, L. Hou, X. He, D. He, S. Chan, and M. Guizani, ``Offloading time optimization via markov decision process in mobile-edge computing," {\emph {IEEE Internet Things J}}., vol. 8, no. 4, pp. 2483--2493, Feb. 2021.
\bibitem{10}
T. Q. Dinh, J. Tang, Q. D. La, and T. Q. S. Quek, ``Offloading in mobile edge computing: Task allocation and computational frequency scaling," {\emph {IEEE Trans}}. {\emph {Commun}}., vol. 65, no. 8, pp. 3571--3584, Aug. 2017.

\bibitem{12} 
G. Gui, H. Huang, Y. Song, and H. Sari, ``Deep learning for an effective nonorthogonal multiple access scheme,'' \emph{IEEE Trans. Veh. Technol.}, vol.~67, no.~9, pp.~8440--8450, Jun. 2018. 
\bibitem{13}
W. Lee, M. Kim, and D. -H. Cho, ``Deep power control: Transmit power control scheme based on convolutional neural network,'' \emph{IEEE Commun. Lett.}, vol.~22, no.~6, pp.~1276--1279, Jun. 2018.
\bibitem{14}
S. Zhu, W. Xu, L. Fan, K. Wang, and G. K. Karagiannidis, ``A novel cross entropy approach for offloading learning in mobile edge computing," {\emph {IEEE Wireless Commun}}. {\emph {Lett}}., vol. 9, no. 3, pp. 402--405, Mar. 2020.
\bibitem{15}
U. Saleem, Y. Liu, S. Jangsher, X. Tao, and Y. Li, ``Latency minimization for D2D-enabled partial computation offloading in mobile edge computing,'' {\emph {IEEE Trans. Veh. Technol.}}, vol.~69, no.~4, pp.~4472--4486, Apr. 2020.

\bibitem{17}
L. Liang, H. Ye, G. Yu, and G. Y. Li, ``Deep-learning-based wireless resource allocation with application to vehicular networks," {\emph {Proc. IEEE}}, vol. 108, no. 2, pp. 341--356, Feb. 2020.

\bibitem{16}
L. Huang, S. Bi, and Y. -J. A. Zhang, ``Deep reinforcement learning for online computation offloading in wireless powered mobile-edge computing networks,'' \emph{IEEE Trans. Mobile Comput.}, vol.~19, no.~11, pp.~2581--2593, Nov. 2020.
\end{thebibliography}
\end{document}